\documentclass[aps,twocolumn,superscriptaddress,floatfix]{revtex4}
\usepackage[dvips]{graphicx}
\usepackage{latexsym}
\usepackage{amsmath}
\usepackage{amsfonts}
\usepackage{amssymb}
\usepackage{bm}
\usepackage{color}
\usepackage{txfonts}
\begin{document}
	\newcommand{\fig}[2]{\includegraphics[width=#1]{#2}}
	\newcommand{\la}{{\langle}}
	\newcommand{\ra}{{\rangle}}
	\newcommand{\dg}{{\dagger}}
	\newcommand{\upa}{{\uparrow}}
	\newcommand{\dna}{{\downarrow}}
	\newcommand{\ab}{{\alpha\beta}}
	\newcommand{\ias}{{i\alpha\sigma}}
	\newcommand{\ibs}{{i\beta\sigma}}
	\newcommand{\hH}{\hat{H}}
	\newcommand{\hn}{\hat{n}}
	\newcommand{\hc}{{\hat{\chi}}}
	\newcommand{\hU}{{\hat{U}}}
	\newcommand{\hV}{{\hat{V}}}
	\newcommand{\br}{{\bf r}}
	\newcommand{\bk}{{{\bf k}}}
	\newcommand{\bq}{{{\bf q}}}
	\def\gsim{~\rlap{$>$}{\lower 1.0ex\hbox{$\sim$}}}
	\setlength{\unitlength}{1mm}
	\newcommand{{\vhf}}{\chi^\text{v}_f}
	\newcommand{{\vhd}}{\chi^\text{v}_d}
	\newcommand{{\vpd}}{\Delta^\text{v}_d}
	\newcommand{{\ved}}{\epsilon^\text{v}_d}
	\newcommand{{\vved}}{\varepsilon^\text{v}_d}
	\newcommand{{\tr}}{{\rm tr}}
	\newcommand{\pprl}{Phys. Rev. Lett. \ }
	\newcommand{\pprb}{Phys. Rev. {B}}

\title {Atomic line defects and topological superconductivity in unconventional superconductors}

\author{Yi Zhang}
\affiliation{Kavli Institute of Theoretical Sciences, University of Chinese Academy of Sciences,
	Beijing, 100190, China}
\author{Kun Jiang}
\email{jiangkun@iphy.ac.cn}
\affiliation{Beijing National Laboratory for Condensed Matter Physics and Institute of Physics,
	Chinese Academy of Sciences, Beijing 100190, China}
\author{Fuchun Zhang}
\affiliation{Kavli Institute of Theoretical Sciences, University of Chinese Academy of Sciences,
Beijing, 100190, China}
\affiliation{Chinese Academy of Sciences Center for Excellence in Topological Quantum Computation,
	University of Chinese Academy of Sciences, Beijing 100190, China}
\author{Jian Wang}
\affiliation{International Center for Quantum Materials, School of Physics, Peking University, Beijing 100871, China}
\affiliation{Chinese Academy of Sciences Center for Excellence in Topological Quantum Computation,
	University of Chinese Academy of Sciences, Beijing 100190, China}
\author{Ziqiang Wang}
\email{wangzi@bc.edu}
\affiliation{Department of Physics, Boston College, Chestnut Hill, MA 02467, USA}

\date{\today}

\begin{abstract}
Topological superconductors (TSCs) are correlated quantum states with simultaneous off-diagonal long-range order and nontrivial topological invariants. They produce gapless or zero energy boundary excitations, including Majorana zero modes and chiral Majorana edge states with topologically protected phase coherence essential for fault-tolerant quantum computing. Candidate TSCs are very rare in nature. Here, we propose a novel route toward emergent
quasi-one-dimensional (1D) TSCs in naturally embedded quantum structures such as atomic line defects in unconventional spin-singlet $s$-wave and $d$-wave superconductors. We show that inversion symmetry breaking and charge transfer due to the missing atoms lead to the occupation of incipient impurity bands and mixed parity spin singlet and triplet Cooper pairing of neighboring electrons traversing the line defect. Nontrivial topological invariants arise and occupy a large part of the parameter space, including the time reversal symmetry breaking Zeeman coupling due to applied magnetic field or defect-induced magnetism, creating TSCs in different topological classes with robust Majorana zero modes at both ends of the line defect. Beyond providing a novel mechanism for the recent discovery of zero-energy bound states at both ends of an atomic line defect in monolayer Fe(Te,Se) superconductors, the findings pave the way for new material realizations of the simplest and most robust 1D TSCs using embedded quantum structures in unconventional superconductors with large pairing energy gaps and high transition temperatures.

\end{abstract}
\maketitle

\section{Introduction}

Superfluid and superconductors are fundamental quantum states exhibiting off-diagonal long-range order \cite{cnyang,anderson}. Such ordered states can be further classified by topologically distinct invariants \cite{altland,class}. Topological superconductors (TSCs) have simultaneous off-diagonal long-range order and nontrivial topological invariants and host topological boundary and defect excitations that are both fundamental and useful for quantum device making \cite{xlqi,read,kitaev,ivanov,nayak}.
The simplest model for a TSC is the Kitaev chain of spinless (single-spin) fermions with $p$-wave pairing \cite{kitaev}. The model does not have time-reversal invariance (TRI) and belongs to the topological class $BDI$ characterized by a nontrivial topological $Z$ invariant \cite{altland,class,tewari-sau}. There are two degenerate zero-energy bound states, i.e. Majorana zero modes (MZMs), spatially localized at each end of the 1D TSC. For realistic spin-${1\over2}$ electrons, theoretical models have been proposed that combine Rashba spin-orbit coupling (SOC), $s$-wave superconductivity, and magnetic Zeeman coupling together to effectively generate such an 1D odd-parity TSC \cite{sau,lutchyn,yuval}. Experimental realizations using hybrid systems of Rashba nanowires proximity coupled to conventional superconductors have made advances \cite{kouwenhoven,patrick,kouwenhoven18,dasarma_rew,lutchyn_rew,hqxu,shtrikman,marcus,marcus16,marcus17,xli,frolov}, but it remains controversial whether the TSC has been realized with MZMs localized at both ends of the nanowire.

\begin{figure*}
	\begin{center}
		\fig{7.0in}{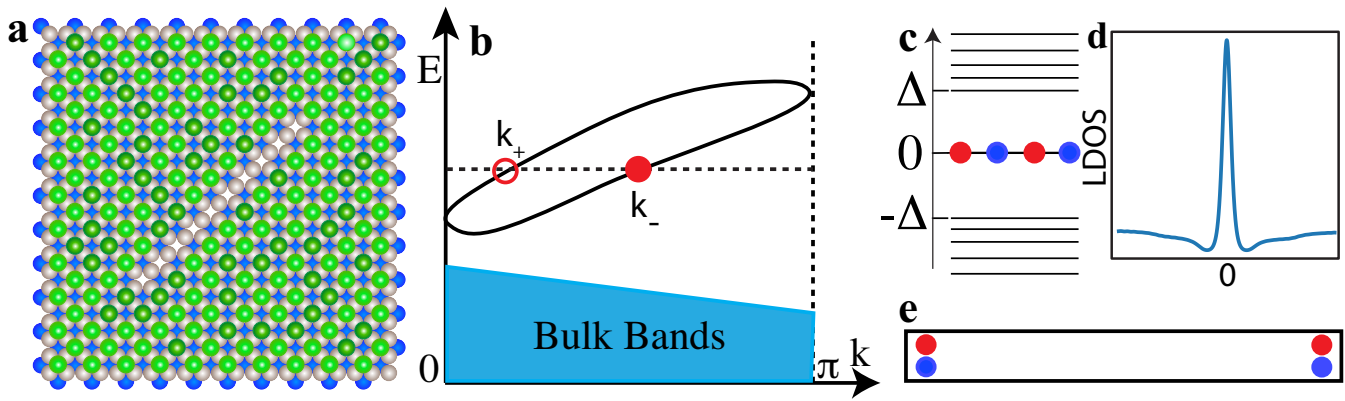}\caption{\textbf{a}, Schematics of monolayer Fe(Te,Se) with an atomic line defect of missing Te/Se atoms along (1,1) direction. Blue and green balls are Te/Se atoms below and above the plane of Fe atoms (silver balls). Random distributed light green balls represent the Se atoms while the dark green balls represent the Te atoms.
			\textbf{b}, Schematics of Rashba split impurity bands.
			The open and solid red circles mark the Fermi points.
			\textbf{c}, Energy spectrum of the TRI TSC
			in a RALD with open boundaries, showing four MZMs (blue and red dots) inside the SC gap. \textbf{d}, Local density of states (LDOS) spectrum at both ends shows a zero-energy conductance peak. \textbf{e}, Blue and red pairs of MZMs form a Kramers doublet,
			localized at both ends of the RALD.
			\label{fig:fig1}}
	\end{center}
	\vskip-0.5cm
\end{figure*}

We report here that quasi-1D TSCs of mixed-parity in multiple topological classes can emerge in naturally embedded quantum structures, such as an atomic line defect, in unconventional spin-singlet superconductors with high transition temperature ($T_c$) and large pairing energy gaps.
The idea was motivated by the recent experimental discovery \cite{jwang} of zero-energy bound states at both ends of an atomic line defect in monolayer high-$T_c$ ($\sim60$K) Fe-based superconductor Fe(Te,Se) grown on SrTiO$_3$ substrates
\cite{fese,mag,xshi,xlpeng}. We will use its atomic structure as an example, but the physics can be more generally applied to other unconventional superconductors. The one-unit-cell Fe(Te,Se) monolayer contains three atomic layers (Fig.~\ref{fig:fig1}{a}). The as-grown atomic line defect corresponds to a line of missing Te/Se atoms on the top layer above the Fe-plane. The missing atoms break the inversion symmetry centered on the Fe atom below, giving rise to a large Rashba SOC $\alpha_R$. To emphasize this property, we will refer to the latter as a Rashba atomic line defect (RALD).

It is important to note that such a quantum structure is very different from the usual crystallography line defects which correspond to edge dislocations or screw dislocations described by the Burgers vectors. The atomic line defects we study here are lines made of point-like defects such as atomic vacancies or adatoms. It is sometimes referred to as the Shockley line defect for its connection to the original work of Shockley on the surface states due to band inversion \cite{shockley}. They appear in metals and narrow-band semiconductors \cite{bandtheorybook}, and more recently in carbon nanotubes \cite{nanotubes} and photonic crystals \cite{photoniccrystals}, as wells as in theoretical model studies of Shockley-Majorana bound states in 2D chiral $p$-wave superconductors \cite{beenakker}. The electronic states in the embedded quantum structure can be described by incipient quasi-1D impurity bands due to the lateral confinement in the bulk superconductor. Moreover, the occupation of the impurity bands is controlled to a large extent by the local electrostatic environment. In the monolayer Fe(Te,Se) \cite{jwang}, the missing Te/Se atoms quench the local $p$-$d$ charge transfer. Since each Fe$^{2+}$ was supposed to transfer two electrons to Te$^{2-}$/Se$^{2-}$, there is an excess of one-electron per unit cell along the RALD. The incipient impurity band can thus be occupied by up to one electron (Fig.~\ref{fig:fig1}{b}). Occupations of more impurity bands may occur more generally \cite{potterlee}. Because of these unique properties of the embedded quantum structure, the coupling of the RALD to the bulk unconventional superconductor can be described microscopically by the coherent processes of hopping, pairing, and spin-orbit coupling, leading to the induced quasi-1D mixed-parity pairing states on the RALD in a systematic and controllable manner. This overcomes the difficulties associated with achieving proximity effect coupling to short coherence length unconventional superconductors using nanowires \cite{law,fan}. Indeed, the results presented here are all obtained from calculations performed directly in the 2D $s$-wave and $d$-wave superconductors with the embedded RALD.

We find that the broken inversion symmetry causes mixed parity pairing \cite{gorkov,nagaosa,qi09,fu10,Fujimoto} in the impurity bands and
the superconducting (SC) states developed in the RALD are TRI quasi-1D TSCs over a large part of the parameter space because of the coherently induced odd-parity spin-triplet pairing. The emerging topological classes and the nature of the zero-energy boundary states turn out to be rich and intriguing, depending on whether a mirror symmetry is present. For unconventional $s$-wave superconductors, a topological crystalline superconductor in the $AIII$ class is realized if the embedded RALD represents a mirror line in the bulk superconductor. The zero-energy bound states at both ends are fermion zero modes in this case, similar to zero-energy Andreev bound states. In contrast, if the mirror symmetry about the embedded RALD is broken, the TRI quasi-1D TSC is in the $DIII$ class characterized by a nontrivial topological $Z_2$ invariant, with a robust Kramers pair of MZMs at each end of the RALD protected by time-reversal symmetry\cite{TRI_reivew}. Since the STM topography in Ref.~\cite{jwang} reveals the disordered Te and Se atoms (dark and light green atoms in Fig.~1a), resulting from the substitutional alloying, that break the mirror symmetry with respect to the line defect, the SC state detected along the atomic line defect in monolayer Fe(Te,Se) may belong to this class. More experiments are necessary and desirable to test if the observed zero-energy bound states at both ends are MZMs.

We also study the RALD embedded in unconventional $d$-wave superconductors such as the high-$T_c$ cuprates. In this case, the $d_{x^2-y^2}$ pairing order parameter is odd under the mirror reflection. We find that the ${\bf d}$-vector of the spin-triplet pairing developed on the impurity bands depends on the direction of the line defect with respect to the nodal line of the bulk $d$-wave superconductor. Remarkably, when the line defect is embedded along the nodal direction, the induced pairing state is a purely parity-odd and mirror-odd spin-triplet SC state.
We show that the resulting TSC is described by two time-reversal partners in class $D$ \cite{sato} with a $Z_2\oplus Z_2$ topological invariant in quasi-1D and Kramers doublets of MZMs at the ends of the RALD. We also study the evolution of these quasi-1D TSCs in response to time-reversal symmetry breaking Zeeman fields due to either the possible incipient magnetism along the RALD or the application of an external magnetic field.

The remaining paper is organized as follows. In section II, we present physical discussions of our findings for the embedded RALD in 2D unconventional $s$-wave and $d$-wave superconductors in terms of simplified 1D effective models. The phase diagram, the topological properties such as the topological invariants and the classification of the TSC, as well as the nature of the zero-energy boundary states will be studied. We then present the 2D model describing the impurity bands along the RALD and its couplings to the bulk superconductor in section III, and discuss the results obtained directly from the 2D unconventional superconductors with the embedded quantum structure. The microscopic realizations, the parameter space, as well as the novelties of the RALD platform will be discussed for both $s$-wave and $d$-wave superconductors. In section IV, the effects of time-reversal symmetry breaking by a vector Zeeman field will be discussed. Summary and outlook will be given in section V.

\section{Effective 1D theory for embedded RALD}

While the microscopic realizations will be derived from the 2D theory later, the general form of the idealized effective 1D model can be written down based on symmetry considerations. For simplicity, we consider a single occupied impurity band $\varepsilon_L(k)$ for the quasiparticles  $f_{k\sigma}$ traversing the line defect, where $k$ denotes the 1D momentum and $\sigma$ the spin. In the Nambu basis $\Psi_k^T=(f_{k\uparrow},f_{k\downarrow},
f_{\bar{k}\uparrow}^\dagger,
f_{\bar{k}\downarrow}^\dagger)$, the Bogoliubov-de Gennes (BdG) Hamiltonian is given by ${\cal H}={1\over2}\sum_k \Psi_k^\dagger H_{1D}(k)\Psi_k$,
\begin{equation}
H_{1D}(k)=\left(
\begin{array}{cccc}
h(k)  & \mathbf{\Delta}(k) \\
\mathbf{\Delta}^\dagger(k) & -h^T(-k)
\end{array}
\right), \label{h}
\end{equation}
where
\begin{equation}
 h(k)=\varepsilon_L(k)\sigma_0+2\alpha_R\sin k\sigma_y+2\alpha_D\sin k\sigma_x,
 \end{equation}
is the normal state Hamiltonian and $\sigma_i$ are the Pauli matrices acting in the spin sector. Here, $\alpha_R$ is the Rashba SOC generated by the broken inversion symmetry, and $\alpha_D$ the Dresselhaus SOC reflecting the possible broken mirror symmetry. When $\alpha_D=0$, the RALD is a mirror line, unless the mirror symmetry is spontaneously broken by the off-diagonal Cooper pairing  $\mathbf{\Delta}(k)$ in Eq.~(\ref{h}). The mixed-parity pairing induced by coupling to the bulk superconductor is given by
\begin{equation}
\mathbf{\Delta}(k)=\Delta_s(k)i\sigma_y + \boldsymbol{d}(k)\cdot{\boldsymbol\sigma}(i\sigma_y),
\end{equation}
where $\Delta_s(k)=\Delta_s(-k)$ denotes the spin-singlet pairing, while $\boldsymbol{d}(-k)=-\boldsymbol{d}(k)$ is the $\mathbf{d}$-vector describing the spin-triplet pairing components\cite{sigrist}.
The 1D BdG Hamiltonian in Eq.~(\ref{h}) has a particle-hole symmetry
$\Theta H_{1D}(k)\Theta^{-1}=-H_{1D}(-k)$ where {$\Theta=\tau_x\sigma_0{\cal K}$} and $\tau_i$ denotes the Pauli matrices acting in the particle-hole sector and ${\cal K}$  the complex conjugation. For a real triplet-pairing $\mathbf{d}$-vector, the effective 1D model is invariant under the time-reversal {${\cal T}H_{1D}(k){\cal T}^{-1}=H_{1D}(-k)$}  where {${\cal T}=i\tau_0\sigma_y{\cal K}$} and ${\cal T}^2=-1$. As a result, $H_{1D}(k)$ also has a chiral symmetry ${\cal C}={\cal T}\Theta$.

Note that the odd-parity $\mathbf{d}$-vector emerges as an intrinsic spin-triplet pairing between the neighboring electrons occupying the impurity bands produced by the embedded RALD.
It does not arise in Rashba nanowires in proximity to $s$-wave superconductors \cite{lutchyn,yuval}. Spin-triplet pairing can be induced in nanowires proximity-coupled to Rashba superconductors \cite{titus,gorkov, dasarma}.
In section III, we will show that the microscopic couplings of the RALD quantum structure to the embedding crystal allow us to determine the directions of the $\mathbf{d}$-vector in 2D unconventional $s$-wave and $d$-wave superconductors.

\subsection{1D model for RALD with ${d_y(k)}$ spin-triplet pairing}

For a RALD embedded in an unconventional $s$-wave SC, we find that the $\boldsymbol{d}$-vector points along the $y$-direction, i.e.
$\boldsymbol{d}(k)=d_y(k){\hat y}$ and $\boldsymbol{d}(k)\cdot{\boldsymbol\sigma}=d_y(k)\sigma_y$. This corresponds to the even combination of odd-parity, equal spin pairing with $d_y(k)=-{i\over2}(\Delta_{\uparrow\uparrow}+\Delta_{\downarrow\downarrow})$. The effective 1D model is therefore,
\begin{eqnarray}
H_{1D}(k)&=&\varepsilon_L(k)\tau_z\sigma_0+2\alpha_R \sin k \  \tau_z \sigma_y+ 2\alpha_D \sin k \ \tau_0\sigma_x
\nonumber \\
&+&\Delta_s(k) \tau_y \sigma_y + \Delta_t(k) \tau_y\sigma_0,
\label{h1d}
\end{eqnarray}
where the triplet pairing is expressed as $\Delta_t(k)=-d_y(k)$.
The impurity band $\varepsilon_L(k)$ splits into two Rashba bands by $\alpha_R$ and $\alpha_D$ with separated pairs of Fermi points at
$\pm k_+$ and $\pm k_-$ as shown schematically in Fig.~1b. They are determined by the condition {$\varepsilon_L(k_\pm){\pm}2(\alpha_R^2+\alpha_D^2)^{1/2}\sin k_\pm=0$}. It turns out that the topological properties of the 1D model in Eq.~(\ref{h1d}) depend crucially on whether a mirror symmetry is present or broken. We thus discuss these two cases separately below.

\subsubsection{RALD as a mirror line: $\alpha_D=0$}

In the absence of the Dresselhaus SOC, i.e. for {$\alpha_D=0$}, the 1D model has a mirror symmetry {${\cal M}_y=-i\tau_0\sigma_y$} and $[H_{1D}(k),{\cal M}_y]=0$.  The embedded RALD thus corresponds to a mirror line in the bulk superconductor, and the effective 1D model describes a topological mirror superconductor \cite{fan13_mirror,sato}. To illustrate this and the nature of the boundary excitations, we block-diagonalize $H_{1D}(k)$ in the mirror eigenbasis where the mirror operator ${\cal M}_y$ is diagonal. Specifically, under the unitary transformation {$U=e^{i\frac{\pi}{4}\tau_0\sigma_x}$}, $U{\cal M}_yU^\dagger={\rm diag} (i,i,-i,-i)$. $H_{1D}(k)$ is simultaneously block diagonal,
\begin{equation}
	UH_{1D}(k)U^\dagger=
	\left(\begin{array}{cc}
	H_A^+ & 0 \\
	0 & H_A^-
	\end{array}\right),
\end{equation}
where $H_A^{\pm}$ are the Hamiltonian in the subspaces with mirror eigenvalues $\pm i$,
\begin{equation}
	H_A^{\pm}=[\varepsilon_L(k)\mp 2\alpha_R \sin k] \tau_z + [\Delta_t(k) \mp \Delta_s(k)]\tau_y.
\end{equation}
Since the mirror operator ${\cal M}_y$ commutes with the particle-hole operator $\Theta$, i.e. $[{\cal M}_y,\Theta]=0$, there is no particle-hole symmetry within each subspace \cite{sato}. Moreover, since there is still a chiral symmetry $\tau_x$, $\{H_A^{\pm},\tau_x\}=0$, each block belongs to the topological class $AIII$ \cite{altland,class}. Thus, the resulting TSC in this case is a topological mirror superconductor characterized by two winding numbers $Z\oplus Z$ \cite{sato,sm}. The zero-energy bound states at the ends of the RALD are fermion zero modes analogous to Andreev bound states, and the mirror symmetry actually prohibits the emergence of MZMs.

\subsubsection{RALD with broken mirror symmetry: $\alpha_D\neq0$}

When the mirror symmetry about the embedded RALD is broken, the effective 1D model must acquire a nonzero Dresselhaus SOC, i.e. $\alpha_D\neq0$. An example is the atomic line defect observed in monolayer Fe(Te,Se) \cite{jwang}. The STM topography reveals the disordered Te and Se distribution, illustrated by the dark and light green atoms in Fig.~1a, that breaks the mirror symmetry with respect to the line defect. Moreover, the zero-bias tunneling conductance  maps indicate that the atomic line defects do not correspond to mirror lines.

For $\alpha_D\neq0$ and however small, the Hamiltonian $H_{1D}$ cannot be diagonalized into reducible mirror eigen blocks. The time-reversal ${\cal T}$, particle-hole $\Theta$, and the chiral symmetry ${\cal C}$ together put the RALD in the class $DIII$ of time-reversal invariant (TRI) superconductors characterized by a topological $Z_2$ invariant (${\cal N}$). A nontrivial ${\cal N}$ corresponds a TRI 1D TSC \cite{class,qi09,qi10,fu10}.
While the latter has been proposed theoretically for nanowires proximity-coupled to iron-based and copper-based superconductors \cite{law,fan}, the role of the mirror symmetry has not been discussed, which will be difficult to control together with the proximity effect superconductivity for such short coherence length superconductors.

The topological invariant ${\cal N}$ can be obtained from the time-reversed pairing function $\delta_{nk}=\langle n,k| {\cal T} \Delta_k^{\dagger} |n,k \rangle$ for each band $n$ \cite{qi10,ardonne} and
${\cal N}=\Pi_{s} [{\rm sgn}(\delta_s)]$, where $s=(n,k_F)$ runs over the Fermi points between $0$ and $\pi$. Thus, an \emph{odd} number of Fermi points with negative time-reversed pairing functions corresponds to a nontrivial $Z_2$ number {(${\cal N}=-1$)} and a TRI TSC. There are four degenerate MZMs in the energy spectrum illustrated in Fig.~1{c}, pairwise localized at both ends of the open chain  in Fig.~1{e}. They give rise to zero-bias conductance peaks at both ends of the RALD as depicted in Fig.~1d.
This TRI but mirror symmetry breaking TSC provides one possible explanation for the SC state detected along the atomic line defect in monolayer Fe(Te,Se), and the zero-energy modes observed at both ends as MZMs. The red and blue pairs of MZMs depicted in Fig.~1e form a Kramers doublet protected by time-reversal symmetry, since the mixing of the MZMs at the opposite ends is exponentially suppressed by the length of the 1D chain \cite{dasarma_split}.
They obey nonabelian braiding statistics and are advantageous for topological quantum computing \cite{xjliu,pgao,kwolm}.

We next obtain the topological phase diagram by calculating the nontrivial topological $Z_2$ invariant ${\cal N}$ for the effective 1D model.
For simplicity, we consider $\varepsilon_L(k)=-2t\cos k-\mu$ and nearest neighbor pairing gap functions $\Delta_s(k)=2\Delta_s\cos k$ and $\Delta_t(k)=2\Delta_p\sin k$. The onsite pairing is ignored since it is suppressed in unconventional superconductors due to the strong local Coulomb repulsion. Evaluating the topological invariant, we obtain
\begin{eqnarray}
{\cal N}={\rm sgn}[(\Delta_{s}\cos k_{+}+\Delta_{p}\sin k_{+})(\Delta_{s}\cos k_{-}-\Delta_{p}\sin k_{-})].
\label{index}
\end{eqnarray}
Calculating ${\cal N}$ results in the topological phase diagram shown in Fig.~2{a} in the $\mu/t-\Delta_p/\Delta_s$ plane, where the TSC (${\cal N}=-1$) occupies a significant part of the phase space. The phase boundaries determined by ${\cal N}$ are further confirmed by performing the Zak phase calculations \cite{ardonne,zak}.
Note that whether a small $\alpha_D$ is present or not does not change the structure of the phase diagram, but does change the symmetry class and the nature of the boundary excitations in the region marked as TRI TSC, as discussed above.
In the limit $\Delta_p/\Delta_s\to\pm\infty$, the 1D chain is always a TRI TSC due to the opposite signs in front of $\Delta_p$ in Eq.~(\ref{index}), which is analytically connected to two Kitaev $p$-wave chains coupled by the Rashba SOC. For finite $\Delta_p/\Delta_s$, the phase boundaries between the TRI TSC and a trivial SC state are given by two gap-closing lines. Along the $\mu=0$ axis in the phase diagram in Fig.~2a, the impurity band is half-filled with $k_\pm$ shifted from ${\pi\over2}$ by an amount $\sim \alpha_R/t$, where the spin-singlet gap function $\Delta_s(k)$ is near its minimum and smaller than the spin-triplet $\Delta_p(k)$ which is near its {maximum}.
This leads to the striking result that the entire line corresponds to TSCs except for the gapless critical point at $\alpha_R/t$. It highlights the robustness of the TSC resulting from the mixed-parity pairing in the RALD.
At nonzero $\mu/t$ and sweeping $\Delta_p/\Delta_s$, the
TSCs are separated by a region of topologically trivial SC state.

\begin{figure}
	\begin{center}
		\fig{3.4in}{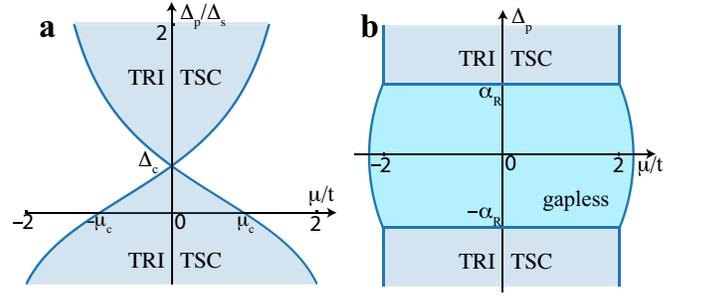}\caption{
{ Topological phase diagram of the effective 1D model. \textbf{a}, RALD embedded in $s$-wave superconductors with mixed parity $s$-wave and $d_y$-vector $p$-wave pairing.
\textbf{b}, Nodal-RALD embedded in $d$-wave superconductors with purely $d_x$-vector $p$-wave pairing. The TRI TSC becomes gapless in the region $-\alpha_R<\Delta_p<\alpha_R$ and bounded by the curved lines from the trivial SC phase.}
			\label{fig:phase_diagram}}
	\end{center}
	\vskip-0.5cm
\end{figure}

\subsection{1D model for RALD with $\boldsymbol{d_x(k)}$ spin-triplet pairing}

Studying an embedded 1D quantum structure in a 2D $d$-wave SC, we find that the topological properties depend on its orientation relative to the nodal direction of the
 pairing gap function. A RALD created along the anti-nodal direction still represents a mirror line, and the results are qualitatively similar to the $s$-wave case discussed above. However, when the atomic line defect is embedded along the nodal lines denoted as the $x$-direction, the $d$-wave pairing potential is {\em odd} under the mirror reflection ${\cal M}_y$.
 Let us refer to this configuration as a nodal-RALD.  Intriguingly, the microscopic couplings between the nodal-RALD and the bulk $d$-wave SC produce an odd-parity spin-triplet pairing with the $\mathbf{d}$-vector pointing along the $x$-direction, i.e. $\boldsymbol{d}(k)=d_x(k){\hat x}$ and $\boldsymbol{d}(k)\cdot{\boldsymbol\sigma}=d_x(k)\sigma_x$. This corresponds to the odd combination of equal spin pairing with $d_x(k)=-{1\over2}(\Delta_{\uparrow\uparrow}-\Delta_{\downarrow\downarrow})$. Moreover, the even parity spin-singlet pairing is forbidden by symmetry.
Consequently, the effective 1D model in Eq.~(\ref{h}) for the nodal-RALD becomes
\begin{equation}
	H_{1D}^{d}(k)=\varepsilon_L(k)\tau_z\sigma_0+2 \alpha_R \sin k\ \tau_z  \sigma_y +\Delta_t(k) \tau_x \sigma_z,
	\label{h1d_d}
	\end{equation}
where the $\alpha_D$ term is left out for simplicity.
Since the triplet pairing part is odd under the ${\cal M}_y$-symmetry discussed above, i.e. it anticommutes with the mirror $\sigma_y$-operation, $H_{1D}^d(k)$ is invariant under ${\cal M}_y^-=-i\tau_z\sigma_y$, and can therefore be block-diagonalized in the eigenbasis of ${\cal M}_y^-$. Specifically, we find that the unitary transformation,
\begin{equation}
	 U_d=\frac{1}{\sqrt{2}}
	 \left(\begin{array}{cccc}
	 0 & 0 &  i & 1\\
	-i & 1 &  0 & 0\\
	 0 & 0 & -i & 1\\
	 i & 1 &  0 & 0
	\end{array}\right)
	\end{equation}
that diagonalizes ${\cal M}_y^-$ by $U_d{\cal M}_y^-U_d^\dagger={\rm diag} ( i,i,-i,-i)$ simultaneously block-diagonalizes $H_{1D}^d(k)$,
\begin{equation}
	U_d  H_{1D}^d(k)U_d^\dagger=
	\left(\begin{array}{cc}
	H^+_D & 0 \\
	0 & H^-_D
	\end{array}\right).
\end{equation}
The Hamiltonians $H_D^\pm$ in the mirror ${\cal M}_y^-$ subspaces that mix the spin and particle-hole sectors are given by

\begin{equation}
	H^{\pm}_D=-\varepsilon_L(k)\tau_z\mp2\alpha_R \sin k \  \tau_0-\Delta_t(k) \tau_x,
\label{hdpm}
\end{equation}
where $\tau_i$ continues to operate in the particle-hole sector in the transformed basis.
Note that since $\{{\cal M}_y^-,\Theta\}=0$, $H_D^\pm$ in each subspace maintains the particle-hole symmetry \cite{sato}. This can be verified directly, as the particle-hole operator in the transformed basis $\Theta_d=U_d\Theta U_d^{\dagger}=\tau_x s_0{\cal K}$, where $s_i$ denotes the Pauli matrices acting in the mirror subspace, is also block-diagonal. In contrast, the time-reversal operator transforms as ${\cal T}_d=U_d{\cal T}U_d^\dagger=-i \tau_z s_x {\cal K}$, which is not block-diagonal and does not preserve time-reversal symmetry in each block. As a result, the effective 1D model is in the topological class $D\oplus D$ characterized by a topological $Z_2$ (or $Z_2\oplus Z_2$) invariant ${\cal N_D}$  \cite{sato,sm}. For a nontrivial ${\cal N_D}$, the nodal-RALD realizes a TSC with a mirror doublet of MZMs at each end of the line.

The energy dispersions of Eq.~(\ref{hdpm}) can be obtained as
\begin{equation}
	E(k)=\pm\left(\sqrt{\varepsilon^2_L(k)+\Delta_t^2(k)}\pm2\alpha_R \sin k\right).
\end{equation}
For $|\Delta_p|>|\alpha_R|$, the spectrum is fully gapped and the topological invariant is nontrivial with ${\cal N_D}=-1$ when $|\mu|<2t$, giving rise to a TRI topological mirror superconductor marked by the blue region in the phase diagram shown in Fig.~2b with a mirror doublet of MZMs at each end of the nodal-RALD. Increasing the magnitude of the chemical potential $\mu$ causes a gap closing transition at $|\mu|=2t$ and turns the TSC into a trivial strong pairing superconductor for $|\mu|>2t$.
Remarkably, the TSC becomes gapless in the region $|\Delta_p|\le|\alpha_R|$, so long as
$\vert\mu \vert \le \mu_c=2(t^2+\alpha_R^2-\Delta_p^2)^{1/2}$ which correspond to two curved boundary lines to gapped topological trivial SC phase as illustrated in Fig.~2b.
%
The intriguing properties of the gapless phase will be studied in the future.

\section{Quasi-1D TSC from embedded RALD in 2D unconventional superconductors}

We next go beyond the effective 1D model and show that this mechanism applies directly to a 2D unconventional $s$-wave or $d$-wave superconductor embedded with a RALD quantum structure. The monolayer FeTe$_{1-x}$Se$_x$ ,
where the electronic structure as a function of $x$ has been studied by angle-resolved photoemission spectroscopy \cite{xshi,xlpeng}, serves as an experimentally available example of unconventional $s$-wave SCs embedded with such a quantum structure \cite{jwang}.
Near $x\sim0.5$, there are two fully occupied $d_{xz/yz}$ bands and one unoccupied, predominantly $p_z$ band just above the Fermi level near the zone center $\Gamma$ point. The electrostatic potential due to the missing negatively charged Te/Se ions pushes the $p_z$ band to cross the Fermi level to accommodate the excess electrons localized around the RALD. Since the two nearly overlapping electron pockets near the zone corner add an even number of degenerate Fermi points when projected onto the RALD, the condition for the nontrivial topological invariant ${\cal N}$ is not affected by the electron bands. We thus consider a single impurity band around the $\Gamma$ point.
\begin{figure}
	\begin{center}
		\fig{3.4in}{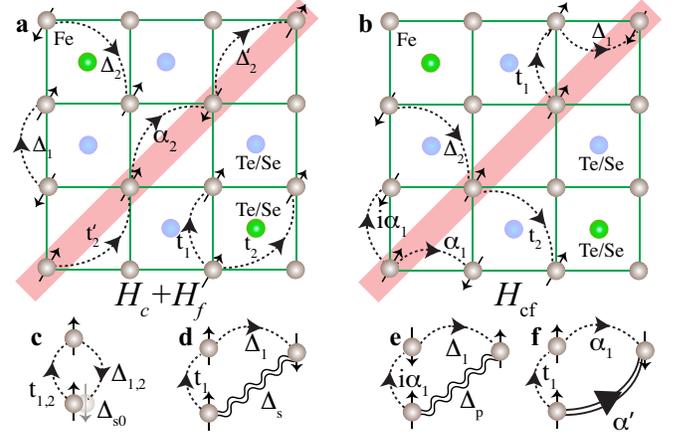}\caption{Schematic illustration of the 2D Hamiltonian for Fe(Te,Se) and the leading dynamic processes generating the mixed-parity pairing along the RALD (highlighted in red). \textbf{a}, Hopping and pairing parameters in the bulk ($H_c$) and along RALD ($H_f$).
			\textbf{b}, Coupling parameters between the bulk and the RALD ($H_{cf}$).
			The 2nd order processes are depicted for generating onsite $s$-wave pairing (\textbf{c}), extended $s$-wave pairing (\textbf{d}), odd-parity $p$-wave pairing (\textbf{e}), and the induced Rashba coupling (\textbf{f}) along the RALD.
			\label{fig:process}}
	\end{center}
	\vskip-0.5cm
\end{figure}

\subsection{2D model of RALD embedded in superconductors}

We use the spinors $c_i=(c_{i\uparrow},c_{i\downarrow})^{T}$ for the electronic states in the bulk and $f_i=(f_{i\uparrow},f_{i\downarrow})^{T}$ for those on the RALD. The 2D model Hamiltonian can be written as $H_{2D}=H_c+H_f+H_{cf}$, with
\begin{eqnarray}
H_c&=&\sum_{ij} -(t_{ij}+\mu_0 \delta_{ij}) c_{i}^\dagger c_{j}+\Delta_{ij} (c_{i}^\dagger i\sigma_yc_{j}^\dagger + h.c.),
\label{hc}
\end{eqnarray}
where $i,j$ run over the 2D lattice of the bulk superconductor except at the line defect. We include the 1st and 2nd nearest neighbor hopping $t_{ij}$ and spin-singlet pairing $\Delta_{ij}$ with amplitudes $t_{1,2}$  and $\Delta_{1,2}$, respectively, as depicted in Fig.~3a. $s$-wave and $d$-wave SCs are distinguished by the sign of $\Delta_{ij}$ under a $C_4$ rotation around a lattice site. The Hamiltonian for the embedded quantum structure $H_f$ is written down on the (Fe) lattice sites along the line defect,
\begin{eqnarray}
	H_f&=&\sum_{\substack{\langle\langle ij\rangle\rangle}} f_{i}^\dagger[-t_{2}^\prime -i\alpha_{2}(\pmb{\sigma}\times \mathbf{\hat{d}}_{ij})_z -i\alpha_D\pmb{\sigma}\cdot\mathbf{\hat{d}}_{ij} + \mu_d \delta_{ij}] f_{j}
\nonumber	\\
	&+&\Delta_{2}^\prime (f_{i}^\dagger i\sigma_yf_{j}^\dagger + h.c.)
	\label{hf}
\end{eqnarray}
where $\mu_d$, $t_2^\prime$, and $\Delta_2^\prime$ are the local potential, the nearest neighbor hopping, and spin-singlet pairing marked in Fig.~3a. Note that $t_2^\prime \ll t_2$ and $\Delta_2^\prime \ll \Delta_2$ due to the missing Te/Se atoms that were otherwise strong facilitators of such the 2nd nearest neighbor processes \cite{yanagi}. In Eq.~(\ref{hf}), $\alpha_2$ and {$\alpha_D$,} are Rashba and Dresselhaus SOCs determined by the unit vector { $\mathbf{\hat{d}}_{ij}=\frac{\mathbf{r}_j-\mathbf{r}_i}{|\mathbf{r}_j-\mathbf{r}_i|}$ connecting sites $i$ and $j$}. For a mirror line $\alpha_D=0$. A small mirror-symmetry breaking $\alpha_D$ is included to account for an embedded RALD without mirror symmetry. This does not produce quantitatively significant changes to the spectrum, but plays an important role in the classification of the TSC and the nature of the zero-energy modes at the ends of the line defect as discussed in the previous section.

The embedded RALD couples to the bulk superconductor according to
\begin{eqnarray}
H_{cf}&=&\sum_{\substack{\langle i,j\rangle }}f_{i}^\dagger[-t_{1}-i\alpha_1(\pmb{\sigma}\times \mathbf{\hat{d}}_{ij})_z]c_{j}+\Delta_{1} \sum_{\substack{\langle i,j\rangle}}f_{i}^\dagger i\sigma_y c_{j}^\dagger \nonumber  \\
&-&t_{2}\sum_{\substack{\langle \langle i,j\rangle \rangle}} f_{i}^\dagger c_{j}+ \Delta_{2}\sum_{\substack{\langle \langle i,j\rangle \rangle}} f_{i}^\dagger i\sigma_y c_{j}^\dagger +h.c.
\label{hcf}
\end{eqnarray}
where, in addition to the couplings by the nearest and second nearest neighbor hopping and pairing terms that appeared in $H_c$ (Eq.~\ref{hc}), the nearest neighbor Rashba SOC $\alpha_1$ is also included as illustrated in Fig.~3b to capture the effects of inversion symmetry breaking.

\subsection{TSC from RALD embedded in $\boldsymbol{s}$-wave superconductors}

We first study the case of $s$-wave SCs, where $\Delta_{1,2}$ are uniform among the nearest and second nearest neighbors, respectively. It is straightforward to show that integrating out the bulk states ($c$ and $c^\dagger$) produces essentially the effective 1D model in Eq.~(\ref{h1d}). The mechanism for dynamically generating the mixed parity pairing state can be understood from the coherent second order processes shown in Figs.~3c-e.
For example, the combination of $\Delta_{1,2}f_{i}(i\sigma_y)c_{j}$ and $t_{1,2}c_{j^\prime}^\dagger f_{i^\prime}$ in Eq.~(\ref{hcf}) induces pairing along the RALD described by $t_\alpha\Delta_\beta G_{j,j^\prime}^cf_i (i\sigma_y) f_{i^\prime}$, where $G_{j,j^\prime}^c=\langle c_jc_{j^\prime}^\dagger\rangle\sim1/\varepsilon_d$ is the equal-time correlator of the bulk states with $\varepsilon_d=\mu_d{-}\mu_0$ the energy separation between the impurity band and the Fermi level.
Processes of this type, as depicted in Figs.~3c-d, give rise to the induced spin-singlet pairing with onsite $s$-wave $\Delta_0\sim t_{1}\Delta_1/\varepsilon_d$ and $t_{2}\Delta_2/\varepsilon_d$,  and nearest neighbor extend $s$-wave $\Delta_s\sim t_{1}\Delta_1/\varepsilon_d$, as well as further neighbor even parity pairing terms along the chain.
Intriguingly, due to the inversion symmetry breaking,
{the SOC combined with $s$-wave pairing between the RALD and the bulk SC produces the odd-parity, spin-triplet pairing in real space as shown in Fig.~3e. For example, $-i\alpha_1 c_{j\downarrow}^\dagger f_{i\uparrow}$ and $\Delta_{1}f_{i^\prime\uparrow}c_{j^\prime\downarrow}$ in Eq.~(\ref{hcf}) generate $-i\alpha_1\Delta_1 G_{j,j^\prime}^{c\downarrow} f_{i^\prime\uparrow}f_{i\uparrow}$.} Processes of this type induce spin-triplet pairing $\Delta_p (e^{i\frac{\pi}{4}}f_{i^\prime\uparrow}f_{i\uparrow}+
e^{-i\frac{\pi}{4}}f_{i^\prime\downarrow}f_{i\downarrow})$ with $\Delta_p\sim \alpha_1\Delta_1/\varepsilon_d$  for the nearest neighbor $p$-wave and further neighbor odd-parity pairing terms involving higher order lattice harmonics. The coherent generation of the odd-parity component in the off-diagonal long-range order is key to produce a robust nontrivial topological invariant to support the TSC.

\begin{figure*}
	\begin{center}
		\fig{7.0in}{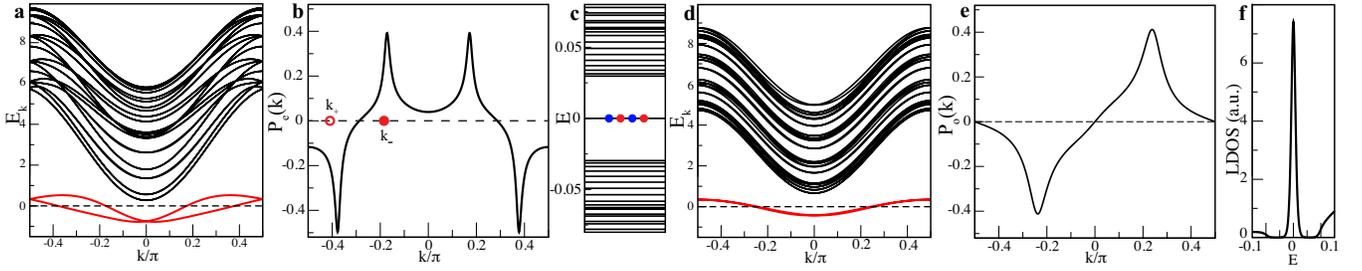}\caption{
\textbf{a-c}, Even-parity dominated quasi-1D TSC. \textbf{a}, Normal state band dispersion plotted in the (1,1) direction in the 2D system with $L_{(1,-1)}=10\sqrt{2}$ and very large $L_D=L_{(1,1)}$ so that $k$ is continuous. The red lines correspond to the incipient impurity bands and the black lines to the bulk bands. \textbf{b}, The expectation value of the induced even parity pairing order parameter $P_e(k)$ in the RALD. The split Fermi points in (\textbf{a}) are marked as $k_+$ and $k_-$ in (\textbf{b}) and enclose within the node of the pairing order parameter $P_e(k^*)=0$. \textbf{c}, Energy spectrum obtained in a finite size system with $L_D= 250\sqrt{2}$, hosting four zero-energy MZMs denoted by blue and red dots. The parameters are $\mu_0=-5.7$, $\mu_d=-0.6$, $t_1=-0.4$, $t_{2}'=0.01$,  $\Delta_1=\Delta_2=0.1$,  $\Delta_2'=0$, $\alpha_1=0$ and $\alpha_2=0.2$. \textbf{d-f}, Odd-parity dominated quasi-1D TSC. \textbf{d}, Normal state band dispersion plotted in the (1,1) direction with the nearly spin-degenerate incipient impurity band in red.
\textbf{e}, The expectation value of the induced odd parity pairing order parameter $P_o(k)$ in the RALD. \textbf{f}, LDOS spectrum at each end of the RALD
in a finite size system with $L_D= 250\sqrt{2}$, showing the zero-energy peak due to the MZMs. The parameters are $\mu_0$=-4.7, $\mu_d$=-0.7, $t_1=0$, $t_{2}'=0.1$, $\Delta_1=0.1$, $\Delta_2=\Delta_2'=0$, $\alpha_1$=0.5 and $\alpha_{2}=0$.
			\label{fig:p-wave}}
	\end{center}
	\vskip-0.5cm
\end{figure*}

\subsubsection{Even-parity dominated quasi-1D TSC}

We now present results directly obtained from diagonalizing the 2D Hamiltonian $H_{2D}$ in Eqs.~(\ref{hc}-\ref{hcf}).
Throughout this section, $t_2$ will be set to unity as the energy unit and {a small $\alpha_D=0.01$ is used unless otherwise noted.} The length of the RALD along the (1,1) direction is denoted by $L_D$, which is embedded in a 2D square lattice of dimensions $L_{(1,1)}$ in the (1,1) and $L_{(1,-1)}$ in the (1,-1) directions under periodic boundary conditions.
The momentum along the RALD is labeled by $k(1,1)$ with $k=n\pi/L_D$ and $n\in[-{L_D\over2},{L_D\over2}]$.
Consider first the case $\alpha_1=0$, where even parity pairing dominates along the RALD.
The normal state energy dispersions
(Fig.~4a) show
the incipient impurity bands (red lines) localized on the line defect. The two split bands due to the Rashba SOC $\alpha_2$ in Eq.~(\ref{hf}) along the line defect cross the Fermi level at $k_\pm$ marked in Fig.~4b. The condition for a nontrivial $Z_2$ invariant ${\cal N}=-1$ in Eq.~(\ref{index}) requires a node in the pairing gap function of the impurity bands to be located in between the two Fermi points.
We thus calculate the induced spin-singlet pairing order parameter $P_{k}=i\sigma_y P_e(k)$ along the embedded line defect. The results show that $P_e(k)$ is a combination of even parity harmonics
and processes a node
between the $k_\pm$ Fermi points ( Fig.~4b), thus realizing a quasi-1D TSC along the RALD in class $DIII$ due to the absence of mirror symmetry. The energy spectrum obtained for the entire 2D sample contains two Kramers doublets of MZMs inside the SC gap as shown in Fig.~4c, which are pairwise localized at each end of the RALD and contribute to the zero-energy peaks in the LDOS.

\subsubsection{Odd-parity dominated quasi-1D TSC}

A crucial part of the topological phase diagram in Fig.~2a is the $p$-wave dominated region. An incipient TRI TSC with dominant odd-parity pairing can indeed emerge from the RALD embedded in bulk $s$-wave superconductors. To this end, we set $\alpha_1=0.5$ and $t_1=\Delta_2=0$, such that the induced even parity pairing is suppressed in the perturbative diagrams in Figs.~3c-d discussed above.
A nearly spin-degenerate impurity band crossing the Fermi level, as shown in Fig.~4d, is obtained when the bare Rashba SOC $\alpha_2$ is set to zero. In the SC state, the induced equal-spin triplet pairing order parameter $P_{k}=\frac{1}{\sqrt{2}}(i\sigma_0+\sigma_z)P_o(k)$, which corresponds to a ${\bf d}$-vector in the direction ${1\over\sqrt{2}}(-1,1,0)$ that is perpendicular to the $(1,1)$ direction of the line defect. In the effective 1D model where the RALD is along the $x$-direction, the ${\bf d}$-vector of the incipient spin-triplet pairing therefore points in the $y$-direction, as described by Eq.~(\ref{h1d}). $P_{k}$ along the embedded line defect can be obtained directly from the 2D calculations.
Fig.~4e shows that $P_o(k)$ is a combination of odd-parity lattice harmonics dominated by the near-neighbor $p$-wave pairing. Indeed, this odd-parity TSC is a realization of two coupled Kitaev $p$-wave chains that respect the time-reversal symmetry. As discussed in section II, in the presence of broken mirror symmetry, the quasi-1D TSC is in class $DIII$.
The energy spectrum of the entire 2D sample contains two Kramers pairs of MZMs inside the SC gap. In real space, they are pairwise localized at each end of the RALD and manifest in the zero-energy peaks in the LDOS as shown in Fig.~4f.

\subsubsection{General quasi-1D TSC of mixed parity}

In general cases relevant to real materials such as Fe(Te,Se),  the SC state developed in the RALD with nonzero $\alpha_1$, $t_{1,2}$, and $\Delta_{1,2}$ is in the regime of mixed parity. We find that the TRI  quasi-1D TSC emerges over a wide region in the phase space.
Fig.~5a displays the normal state band dispersions obtained in the 2D lattice calculation for a general set of parameters with a partially filled bulk band, showing multiple impurity bands crossing the Fermi level.
We calculate the dynamically generated mixed parity pairing order parameter $P_{k}=i\sigma_y P_e(k)+{1\over\sqrt{2}} (i\sigma_0+\sigma_z)P_o(k)$ along the line defect. The results plotted in  Fig.~5b show that the even ($P_e$) and odd ($P_o$) parity components are of comparable magnitudes, composed of mixtures of $s$-wave and $p$-wave and their higher order harmonics. Since the line defect is not a mirror line, the TRI quasi-1D superconductor is in class $DIII$ characterized by the $Z_2$ invariant ${\cal N}$ in Eq.~(\ref{index}). Remarkably, close to the momenta of the Rashba-split Fermi points, the odd-parity pairing is at its maximum and dominates over the even-parity pairing amplitude, resulting in a nontrivial ${\cal N}=-1$ from Eq.~(\ref{index}) and a TRI quasi-1D TSC along the RALD. The topological nature of the superconductor is also confirmed by the Zak phase calculation~\cite{zak}.
The energy spectrum of the 2D superconductor shows four MZMs inside the SC gap. The calculated {LDOS} spectra along the entire line defect shown in Fig.~5c displays two zero-energy peaks localized near both ends where the two Kramers doublets of MZMs reside, and a rather clean SC spectrum without in-gap states in the middle region of the RALD. These results are in good agreement with the evolution of the STM tunneling conductance spectra along the atomic line defect in monolayer Fe(Te,Se) \cite{jwang}.

\begin{figure*}
	\begin{center}
		\fig{7.0in}{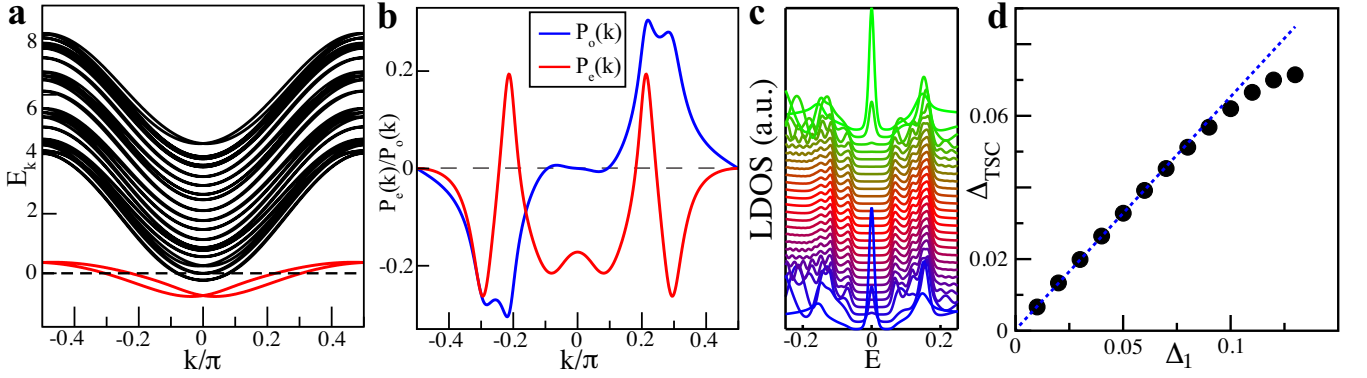}\caption{Quasi-1D TSC in the general region of mixed parity. \textbf{a}, Normal state band dispersions in the (1,1) direction in the 2D system with $L_{(1,-1)}=10\sqrt{2}$ and large $L_D=L_{(1,1)}$ so that $k$ is continuous.  The red lines correspond to the impurity bands and the black lines to the bulk bands.
			\textbf{b}, Expectation values of the pairing order parameters with both even parity ($P_e$) and odd parity ($P_o$) components along the RALD. \textbf{c}, The LDOS spectra along the line-cut through the RALD from one end (bottom) to the other end (top) taken at equal spatial distance of $10\sqrt{2}$, with the exception that the second curve from the bottom and top corresponds to the points $5\sqrt{2}$ from the two ends. The RALD is embedded in a finite size system along (1,1) direction with $L_D= 250\sqrt{2}$, $L_{(1,1)}=500\sqrt{2}$, and $L_{(1,-1)}=10\sqrt{2}$.
			{\textbf{d},} The evolution of the topological gap $\Delta_{TSC}$ along the RALD as a function of the bulk SC gap $\Delta_1$ while the other parameters are fixed.
			Parameters used are: $\mu_0$=-4.0, $\mu_d$=-0.8, $t_{2}'=0.1$, $t_1=0.1$, $\Delta_1=0.1$, $\Delta_2=\Delta_2'=0$, $\alpha_1$=0.5, and  $\alpha_{2}=0$.
			\label{fig:ps-wave}}
	\end{center}
	\vskip-0.5cm
\end{figure*}

One important advantage for materializing the quasi-1D TSC with an intrinsic quantum structure embedded in unconventional superconductors is that the incipient TSC and the MZMs can be protected by a large SC gap and operate at higher temperatures. For example, the perturbation analysis discussed above indicates that the nearest-neighbor $s$-wave and $p$-wave pairing gaps follow $\Delta_s\sim t_{1} \Delta_1/\varepsilon_d$ and $\Delta_p\sim \alpha_1 \Delta_1/\varepsilon_d$.
Since both pairing gaps scale with the bulk SC gap $\Delta_1$, the topological gap along the line defect should be proportional to the SC gap of the unconventional superconductor. However, because the induced pairing functions along the line defect are complicated and involve further neighbors in both the even and odd parity channels (Fig.~5b), it is
much more reliable to calculate the topological gap $\Delta_{TSC}$ of the impurity bands directly from the 2D model with the embedded line defect. In Fig.~5d,
the extracted $\Delta_{TSC}$ along the RALD from the lowest excitation energy of the 2D system is plotted as a function of the bulk SC gap  $\Delta_1$, while keeping the other parameters unchanged as in Figs. 5a-c. It shows a good linear relationship $\Delta_{TSC}\sim 0.65\Delta_1$
and demonstrates that the incipient topological gap can be a significant fraction of the bulk SC gap in the unconventional superconductor, provided that the Rashba SOC is strong. For Fe(Te,Se), recent DFT calculations provide an estimate of the Rashba SOC $\alpha_1\sim60$meV ($\sim0.16$ eV$\cdot\AA$ in the standard unit) \cite{wu_soc}. First principle calculations also predict that the bulk bandwidth for the $p_z$ band is around $1$eV \cite{wu_t}, which is further reduced by the correlation effects \cite{watson}. This provides an estimate of the hopping parameter $t_2\sim125$ meV and the ratio $\alpha_1/t_2\sim0.5$. Thus, the Rashba SOC is in the range of the parameters used in Fig.~5 and strong enough to produce a sizable incipient topological gap along the RALD as a significant fraction of the large bulk SC gaps ($11$meV and $18$meV) observed in monolayer Fe(Te,Se) \cite{jwang}.
%

To explore the robustness of the quasi-1D TSC with Majorana zero-energy end states, we have also studied non-straight RALDs composed of continuous zigzag segments embedded in the unconventional $s$-wave superconductor as shown in {Figs.~6a and 6c}. Here, we set the Dresselhaus SOC $\alpha_D=0$ since the zigzag shape of the line defect already breaks the mirror symmetry. For both types of non-straight RALDs, our 2D calculations find a Kramers doublet of MZMs localized at each end that gives rise to the zero-bias peaks in the LDOS shown in {Figs.~6b and 6d}. The localization lengths of the zero modes are very short and on the order of a lattice constant wherein the local environment near the ends of the non-straight line defects are essentially the same, leading to the nearly identical LDOS in Figs.~6b and 6d. This demonstrates that the zero energy modes are robust against the changes in the shape of the line defect as long as the quasi-1D TSC developed along the line defect remains stable.

\begin{figure}
	\begin{center}
		\fig{3.4in}{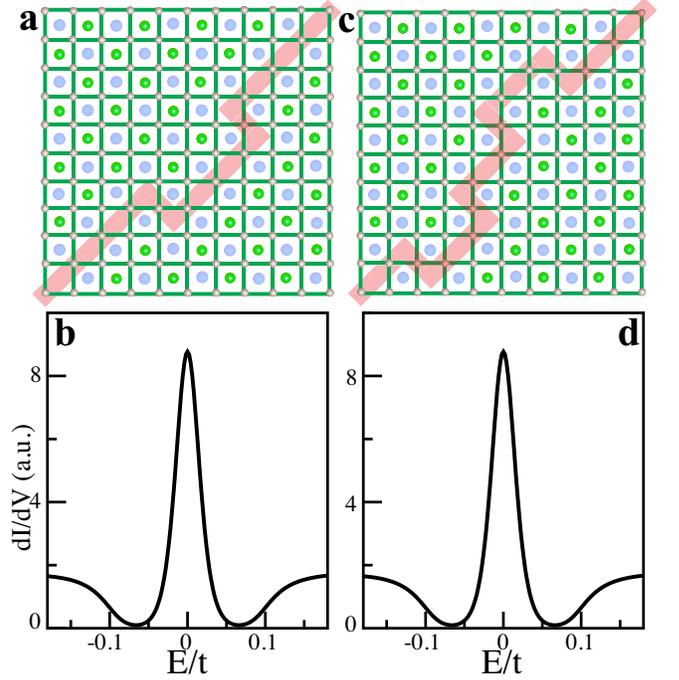}\caption{Quasi-1D TSC along non-straight RALD. \textbf{a} and \textbf{c}: Schematics of two different zigzag line defects. The total length of the system along the (1,1) direction is 500$\sqrt{2}$. The lengths of the segments are, from lower left to upper right, 104$\sqrt{2}$, {$\sqrt{2}$}, 41$\sqrt{2}$, {$\sqrt{2}$}, and 104$\sqrt{2}$ for \textbf{a}; and 54$\sqrt{2}$, {$\sqrt{2}$}, 30$\sqrt{2}$, {$\sqrt{2}$}, 30$\sqrt{2}$, {$\sqrt{2}$}, 30$\sqrt{2}$, {$\sqrt{2}$}, and 105$\sqrt{2}$ for \textbf{c}. \textbf{b} and \textbf{d}, The corresponding LDOS at both ends of the non-straight line defects, showing robust zero-bias peaks due to the zero-energy end states.
The parameters used are: $\mu_0$=-4.0, $\mu_d$=-0.8, $t_{2}'=0.1$, $t_1=0.1$, $\Delta_1=0.1$, $\Delta_2=\Delta_2'=0$, $\alpha_1$=0.5, and  $\alpha_{2}=0$.
        \label{fig:zigzag}}
	\end{center}
	\vskip-0.5cm
\end{figure}

\subsection{TSC from RALD embedded in $\boldsymbol{d}$-wave superconductors}

Due to the electron-electron correlations, there is another important class of unconventional superconductors, the $d$-wave superconductors, discovered in the high-$T_c$ cuprates and certain heavy-fermion superconductors. In this section, we study embedded 1D quantum structures such as the RALD in unconventional $d$-wave superconductors on the 2D square lattice. This system continues to be described by the Hamiltonian $H_{2D}$ given in Eqs.~(\ref{hc},\ref{hf},\ref{hcf}), with a sign-changing nearest neighbor pairing field $\Delta_{ij}$ under the $C_4$ rotation. For simplicity, the second nearest neighbor pairing, Rashba SOC, and hopping amplitudes will be set to zero, i.e. $\Delta_2=\Delta_2^\prime=\alpha_2=t_2=t_2^\prime=0$.
Because the $d_{x^2-y^2}$-wave pairing order parameter changes signs under the $C_4$ lattice rotation, the nature of the incipient TSC depends intriguingly on the direction of the line defect with respect to the nodal line of the bulk $d$-wave superconductor. We thus discuss these cases separately.

\subsubsection{RALD embedded along antinodal directions}

\begin{figure}
	\begin{center}
		\fig{3.4in}{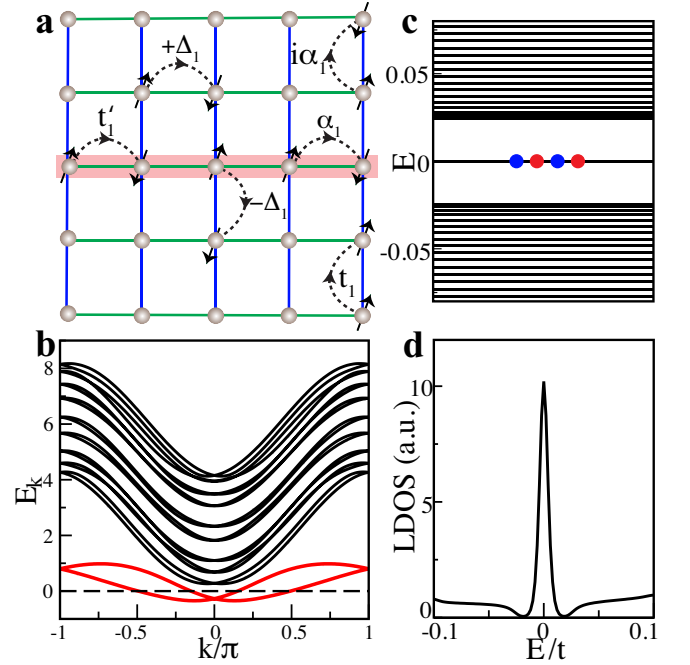}\caption{ \textbf{a}, Schematic illustration of a RALD (shaded region) embedded in the anti-nodal direction of a 2D $d$-wave superconductor. The green and blue bonds correspond to positive and negative pairings respectively. \textbf{b}, Normal state band dispersions. The momentum is along the $x$-direction of the line defect with $L_D=L_{(1,0)}=1000$ and $L_{(0,1)}=10$. The red lines correspond to the incipient impurity bands split by Rashba SOC. The black lines correspond to the bulk bands. \textbf{c}, The energy spectrum of the 2D superconductor with $L_D=500$, extending from (250,5) to (750,5). The four zero-energy states are represented by the blue and red dots. \textbf{d}, The LDOS spectrum at either end of the line defect. A temperature broadening ($T=0.0025$) is included. Parameters used are $t_1$=-1, $t_1^{\prime}$=-0.1, $\alpha_1$=0.2, $\alpha_D$=0.01, $\Delta_1$=0.1, $\mu_0$=-4.2 and $\mu_d$=-1.0.
			\label{fig:dwave-x}}
	\end{center}
	\vskip-0.5cm
\end{figure}

Consider the line defect embedded along the $x$ direction, which is the antinodal direction of the nearest neighbor $d$-wave pairing with $\Delta_x=-\Delta_y=\Delta_1$, as shown in Fig.~\ref{fig:dwave-x}a. As in the $s$-wave case, the RALD is characterized by a local electrostatic potential $\varepsilon_d$ and an inversion symmetry breaking induced Rashba SOC $\alpha_1$ originating, for example, from a line of missing off-plane atoms such as the apical oxygens or the La atoms in the cuprates. In the numerical calculations, the nearest-neighbor hopping coupling between the line defect and the bulk superconductor ($t_{1d}$) is allowed to be smaller than the corresponding $t_1$ in the bulk to model the localized electronic structure of the embedded quantum structure. The normal state band dispersions obtained in the 2D model have a pair of incipient Rashba split impurity bands cross the Fermi level, as shown in Fig.~\ref{fig:dwave-x}b.

The microscopic mechanism for generating the quasi-1D SC state by the coherent couplings of the line defect to the bulk superconductor is similar to the $s$-wave case discussed above and gives rise to the important odd-parity, spin triplet pairing component. For the RALD aligned in the antinodal direction, it can be seen from Fig.~\ref{fig:dwave-x}a that the pairing potential maintain the mirror symmetry ${\cal M}_y$ about the line defect. As a result, the induced pairing state is of mixed parity described by $\Delta_s(k)i\sigma_y + \boldsymbol{d}(k)\cdot{\boldsymbol\sigma}(i\sigma_y)$ with the spin-triplet pairing ${\bf d}$-vector pointing in the $y$-direction, i.e.
$\boldsymbol{d}(k)\cdot{\boldsymbol\sigma}=d_y(k)\sigma_y$, leading to the same effective 1D model given in Eq.~(\ref{h1d}), as in the $s$-wave case.
%
The energy spectrum in Fig.~\ref{fig:dwave-x}c
obtained for the 2D superconductor shows that a TRI quai-1D TSC emerges with four zero-energy states pairwise localized at the ends of the embedded RALD. Similar to the $s$-wave case, in the presence of mirror symmetry, this would be a topological mirror superconductor in class $AIII\oplus AIII$ with fermion zero-modes at both ends. When the mirror symmetry is broken (such as by a nonzero $\alpha_D$), the TSC is in class $DIII$ with a Kramers doublet of MZMs localized at each end of the RALD, where the tunneling spectrum exhibits the zero-energy conductance peak as shown in Fig.~\ref{fig:dwave-x}d.

\begin{figure*}
	\begin{center}
		\fig{7.0in}{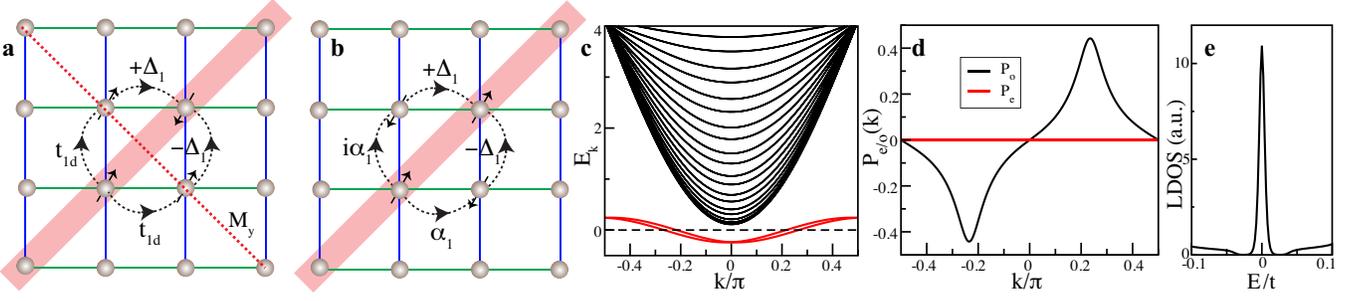}\caption{\textbf{a-b}, Schematic illustration of a nodal RALD  embedded in the nodal direction of a $d$-wave superconductor. The mirror operation $M_y$ is indicated by the red dash line in \textbf{a}. The pairing amplitudes on  the green and blue lattice bonds have opposite signs. The two processes for spin-singlet pairing along the RALD cancel  in \textbf{a} and the two processes shown in \textbf{b} contribute to spin-triplet pairing along the RALD.
			\textbf{c}, Band dispersions for the normal state in the (1,1) direction in the 2D system with $L_{(1,-1)}=10\sqrt{2}$ and very large $L_D=L_{(1,1)}$ so that $k$ is continuous.
			 The red lines correspond to the dispersions of the incipient impurity bands split by Rashba SOC. The black lines are the dispersions of the bulk bands. \textbf{d}, Expectation values of the induced pairing order parameters in the RALD for the even parity $P_e(k)$ and odd parity $P_o(k)$, where $P_e(k)$ is zero due to symmetry.
			 \textbf{e}, The LDOS spectrum at either end of the line defect with a temperature broadening {T=0.0025}.
			 Parameters used are: $\mu_0$=-4.1, $\mu_d$=-0.3, $t_{1}=-1$, $t_2$=0, $t_2'$=-0.1,    $t_{1d}$=-0.05, $\Delta_1=0.1$, $\Delta_2=\Delta_2'=0$, $\alpha_1$=0.5, and  $\alpha_{2}=0$.
			\label{fig:dwave}}
	\end{center}
	\vskip-0.5cm
\end{figure*}

\subsubsection{RALD embedded along nodal directions}

The more intriguing case is when the quantum structure is embedded along the nodal direction of the bulk $d$-wave superconductor. Such a RALD aligned with the $(1,1)$ direction is depicted in Fig.~\ref{fig:dwave}a-b and produces the Rashba-split impurity bands shown in Fig.~\ref{fig:dwave}c. The important point to notice is that the $d$-wave pairing field is odd under the mirror operation {${\cal M}_y$} marked in Fig.~\ref{fig:dwave}a. In contrast to the antinodal and the $s$-wave cases, the induced pairing along the line defect cannot sustain an $s$-wave component which is even under {${\cal M}_y$}. Intuitively, this can be seen from the second order processes shown in Fig.~\ref{fig:dwave}a, where the contributions from the two paths involved in producing a nearest neighbor pairing along the RALD are out of phase and cancel out in the spin-singlet channel. However, odd-parity, spin triplet pairing that is odd under the mirror {${\cal M}_y$} can still be induced along the line defect by the combined processes of nearest neighbor hopping and Rashba SOC as illustrated in Fig.~\ref{fig:dwave}b. Therefore, the nodal RALD in the $d$-wave setup gives a unique platform for generating purely spin-triplet quasi-1D SCs and can be further tested by experiments.

Carrying out the 2D calculations, we find that the only incipient pairing induced coherently on the impurity bands is the equal-spin triplet pairing order parameter $P_{k}={1\over\sqrt{2}}(i\sigma_0-\sigma_z) {P_o(k)}$. The minus sign indicates that the ${\bf d}$-vector of the spin-triplet pairing is different from that in the RALD embedded in $s$-wave or along the antinodal direction in $d$-wave superconductors. The ${\bf d}$-vector is thus given by ${1\over\sqrt{2}}(1,1)$, pointing along the direction of the RALD. If we rotate the coordinates so that the RALD aligns with the $x$-direction, the triplet pairing ${\bf d}$-vector is then in the $x$-direction, i.e.
$\boldsymbol{d}(k)\cdot{\boldsymbol\sigma}=d_x(k)\sigma_x$, giving rise to the odd-combination of the equal spin pairing described by the effective 1D model in Eq.~(\ref{h1d_d}). Fig.~\ref{fig:dwave}d displays the vanishing even-parity $P_e$ and the robust odd-parity $P_o$ components of the pairing order parameter obtained from diagonalizing the 2D model with the embedded quantum structure.
From the study of the effective 1D model in Eq.~(\ref{h1d_d}), a quasi-1D topological crystalline superconductor is clearly realized in class $D\oplus D$ characterized by the nontrivial $Z_2\oplus Z_2$ invariant, with a mirror doublet of MZMs pairwise localized at each end of the embedded nodal RALD, where the tunneling spectrum exhibits the zero-energy conductance peak shown in Fig.~\ref{fig:dwave}e.


\section{Zeeman effect and time-reversal symmetry breaking TSC}

The discussions thus far have focused on the situation with time-reversal symmetry. In this section, we study the response of the TSC developed in the RALD along the $x$-direction to general time-reversal symmetry breaking Zeeman fields,
\begin{equation}
H_Z=h_x\sigma_x\tau_z+h_y\sigma_y\tau_0
+h_z\sigma_z\tau_z.
\label{hz}
\end{equation}
The reason is two-fold. First, it is possible that the line of missing atoms causes incipient local magnetic order in the neighboring atoms. For example, in the monolayer Fe(Te,Se), the line of missing Te/Se atoms may cause the Fe atoms underneath to become magnetic. The current experiments have not seen evidence for this to happen, but more experiments are necessary \cite{jwang}. Second, the evolution of the SC state with an applied vector external magnetic field ${\bf h}=(h_x,h_y,h_z)$ can be used experimentally to probe the nature of the TSC.

\subsection{RALD embedded in $\boldsymbol{s}$-wave superconductors}

{\em Mirror symmetric RALD} -- We first study the case where the system has perfect mirror symmetry, so that the Dresselhaus SOC vanishes, i.e. $\alpha_D=0$.
Any ${\bf h}\ne0$ in Eq.~(\ref{hz}) breaks the time reversal ${\cal T}$ in the effective 1D model in Eq.~(\ref{h1d}).  For a field in the $(x,z)$ plane, i.e. ${\bf h}_{xz}=(h_x,0,h_z)$, the mirror symmetry ${\cal M}_y$ is also broken, so that $H_{1D}$ cannot be block-diagonalized in the mirror subspace and is thus no longer in class $AIII$. However, the combined operation ${\widetilde{\cal T}}={\cal M}_y{\cal T}={\cal K}$ remains a {hidden} time-reversal symmetry, i.e. ${\widetilde{\cal T}}H_{1D}(k){\widetilde{\cal T}}^{-1}=H_{1D}(-k)$ with ${\cal\widetilde T}^2=+1$, and leads to a hidden chiral symmetry ${\cal\widetilde C}={\cal\widetilde T}\Theta=\sigma_0\tau_x$.
As a result, for exchange field or applied magnetic field in the $(x,z)$-plane, the TSC turns into the $BDI$ class characterized by an integer topological invariant $Z$ \cite{tewari-sau,tewari}. The zero-energy modes remain stable and become two pairs of MZMs localized at the ends of the mirror line RALD.  
One arrives at the same conclusion for the 2D embedded RALD governed by $H_{2D}$ when ${\bf h}_{xz}\neq0$, with four MZMs protected by the hidden chiral symmetry as shown by direct 2D calculations shown in Fig.~9a.
For large enough ${\bf h}_{xz}$, one of the Rashba-split bands (Figs.~4a) is pushed to strong pairing outside the Fermi level.
The quasi-1D TSC thus exhibits a single MZM at each end of the RALD as shown in Fig.~9a. This is analogous to the TSCs proposed for Rashba quantum wires \cite{sau}, magnetic Fe-chains \cite{yazdani,titus,jian} proximity coupled to s-wave superconductors in the presence of Zeeman fields, as well as the TSCs realized in multichannel Shiba chains \cite{junhua,gian}. The quantization of the tunneling conductance at the ends changes from ${4e^2\over h}$ to ${2e^2\over h}$. The evolution of the TSC can be studied by applying magnetic field in the plane spanned by the mirror RALD direction and the normal of the 2D SC plane. Interestingly, rotating the magnetic field such that there is a nonzero field component out of this plane, i.e. $h_y\ne0$, removes the hidden  ${\cal \widetilde T}$ and thus breaks the chiral symmetry ${\cal \widetilde C}$, leaving the system in class $A$ which is topologically trivial in 1D \cite{class}. Thus, a magnetic field or an exchange field component $h_y\ne0$ destroys the TSC together with the zero modes at the ends of the RALD, as shown by the results obtained directed in the 2D superconductor plotted in Fig.~9b.

\begin{figure}
	\begin{center}
		\fig{3.4in}{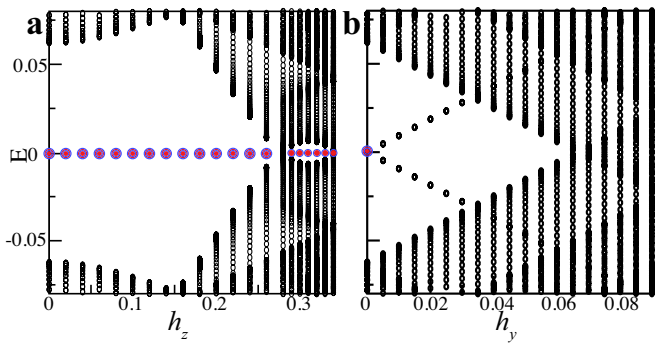}\caption{\textbf{a}, Energy spectrum of the 2D $s$-wave superconductor with an embedded mirror-RALD as a function of the Zeeman field $h_{z}$.
			The SC gap closes and reopens at $h_z^c\simeq0.28$ with the concomitant reduction in the number of MZMs from four to two (red and blue dots), signaling the transition of the TSC in class $BDI$ with the reduction of the topological invariant from $Z=2$ to $Z=1$.
			\textbf{b}, Same as in \textbf{a} and with the same parameters, but the Zeeman field $h_y$ is perpendicular to the mirror-RALD. The zero modes split immediately as $h_y$ is turned on, as the TSC becomes a topologically trivial superconductor.}
			\label{fig:zeeman}
	\end{center}
	\vskip-0.5cm
\end{figure}

{\em Mirror symmetry breaking RALD} -- When mirror symmetry is broken, e.g. by a nonzero Dresselhaus SOC $\alpha_D\neq0$ in Eq.~(\ref{h1d}), the time-reversal symmetric TSC is in the $DIII$ class with a Kramers pair of MZMs at each end of the RALD. In this case, the Zeeman field ${\bf h}$ in any direction breaks the time-reversal symmetry completely and no hidden time reversal symmetry can arise due to the broken of mirror symmetry. As a result, the system makes a transition to the topological trivial phase and the unprotected zero-energy Majorana end states are destroyed. As discussed in earlier sections, the mirror symmetry is broken in real materials such as the Fe(Te,Se) due to substitutional alloying of Te and Se atoms as well as the presence of other line defects. If such disorder effects imply that the RALD cannot represent a mirror line, then applying a magnetic field in any direction would destroy the zero-energy bound states observed experimentally at the ends of the RALD. It is thus crucial to carry out these experiments to help determine the nature of the quasi-1D TSC detected in the monolayer Fe(Te,Se) \cite{jwang}. We have also studied the case where the Zeeman field is applied only along the atomic line defect and obtained essentially the same results.

\subsection{Nodal-RALD embedded in $\boldsymbol{d}$-wave superconductors}

Next, we study the Zeeman field response of the TSC developed in the nodal-RALD embedded in unconventional $d$-wave superconductors. In this case, the emergent SC state is described by the mirror-odd, purely spin-triplet pairing with the $\textbf{d}$-vector pointing along the line defect given in Eq.~(\ref{h1d_d}). We have shown that the BdG Hamiltonian can be block-diagonalized in the eigen basis of an effective mirror symmetry ${\cal M}_y^-=-i\sigma_y\tau_z$, where each block describes a TSC in class $D$ with a $Z_2$ topological invariant.

For a Zeeman field in the $y$-direction, the effective mirror symmetry remains since $[h_y\sigma_y\tau_0,{\cal M}_y^-]=0$. The BdG Hamiltonian continues to be block-diagonal in the eigen basis of ${\cal M}_y^-$, with particle-hole symmetry in each block.
Therefore, despite the breaking of time-reversal symmetry by the Zeeman field $h_y$, the system still consists of two sub-blocks (unrelated by time-reversal) and belongs to class $D\oplus D$. The mirror doublet of MZMs at each end of the nodal RALD remains stable.
On the other hand, a Zeeman field $h_{xz}$ in the $xz$-plane breaks the effective ${\cal M}_y^-$ mirror symmetry so that the BdG Hamiltonian is no longer block-diagonal.
The quasi-1D superconductor without time reversal symmetry is thus in class D characterized by a $Z_2$ invariant. We find that the MZMs split immediately upon switching on the Zeeman field  $h_{xz}\neq0$, as the quasi-1D superconductor enters the topologically trivial phase.


\section{Summary and Outlook}

We presented a new route for materializing quasi-1D topological superconductors using naturally embedded quantum structures, such as the Rashba atomic line defects, in spin-singlet unconventional superconductors. The advantage of this platform, besides the high transition temperature and large pairing energy gap, is that the topological superconductivity develops from the coherent quantum processes via the microscopic couplings between the incipient impurity bands and the bulk superconductor. This avoids relying on proximity-effect induced superconductivity that is difficult to achieve and hard to control in unconventional superconductors, and at the same time, brings out the crystalline symmetry for the emergent TSCs to be reckoned with. As a result, several quasi-1D TSCs in distinct topological classes with different zero-energy boundary states can be realized, all involving the odd-parity, spin-triplet pairing order parameter with the ${\textbf{d}}$-vector microscopically determined by direct calculations in the bulk superconductor with the embedded quantum structure.

For a mirror-line RALD embedded in $s$-wave superconductors, a quasi-1D mirror TSC in class $AIII$ arises with zero-energy fermion end states. When the mirror symmetry is broken, the TRI quasi-1D TSC is in class $DIII$ characterized by a nontrivial $Z_2$ invariant, with a Kramers doublet of MZMs localized at each end. The two cases can be distinguished by their different responses to the Zeeman field due to an external magnetic field or an incipient magnetic order. While a nonzero Zeeman field in any direction destroys the mirror-broken, time-reversal symmetry protected TSC in class $DIII$ together with the zero-energy bound states, a Zeeman coupling in the plane spanned by the line defect and the normal direction ($xz$-plane) puts the mirror-line TSC in class $BIII$ with MZMs end states characterized by a nontrivial topological $Z$ invariant. Yet another interesting quasi-1D topological crystalline superconductor in class $D$ emerges when a RALD is embedded in a $d$-wave superconductor along the nodal direction of the $d$-wave gap function. Two pairs of mirror doublet of MZMs emerge at both ends of the line defect, which remain stable under a Zeeman field aligned in the $y$-direction, but are destroyed together with the TSC when the Zeeman field has a nonzero component in the $xz$-plane.

This new mechanism for the emergent quasi-1D TSC provides a possible explanation for the zero-energy bound states discovered at the ends of the atomic line defects in monolayer Fe(Te,Se) high-$T_c$ superconductors. However, the nature of the zero-energy end states, i.e. MZMs versus fermion zero-energy bound states, depends on whether the atomic line defect breaks the mirror symmetry, 
as discussed above. Although we argued that the substitutional Te/Se alloying breaks the mirror symmetry, whether these zero-energy end states are indeed MZMs requires further experimental investigation. Their responses to an applied magnetic field discussed here offer a concrete and amenable experimental test.

There are a few key ingredients that enable this mechanism for the emergent TSCs in embedded quantum structures. At the very basic level, it requires the occupation of an incipient impurity band in the presence of a large Rashba SOC associated with the local inversion symmetry breaking. This is sufficient for producing a quasi-1D SC state of mixed parity by the microscopic couplings of the impurity band states to the bulk spin-singlet unconventional superconductor, without detailed requirements on the shape and length of the 1D quantum structure. The occupation of the impurity band is determined by the local electrostatic environment and plays an important role. The most robust quasi-1D TSC emerges when the impurity band is around half-filling, where the induced spin-triplet pairing dominates over the singlet pairing near the Fermi level and supports a nontrivial topological invariant along the quantum structure. The atomic line defect of Te/Se vacancies in monolayer Fe(Te,Se) films turns out to contain these essential ingredients with an excess of close to one-electron per unit cell occupying an incipient impurity band under a strong Rashba SOC.
We note that Teo and Kane have studied the defect states such as crystalline dislocations in superconductors belonging to different topological classes \cite{teo}. They considered the topological defects (points, lines, and surfaces) as boundaries of the superconductors and classified the emergent gapless defect excitations. What we have shown here is that embedding a lower dimensional system in a higher dimensional topological trivial superconductor can generate a topological superconductor in the embedded quantum structure.
Once a topological superconductor of a specific class emerges in the embedded lower-dimensional superconductor, the nature of the defect and boundary excitations can be classified according to Ref.\cite{teo}.  It may be possible to build a connection between our findings and the classification of topological defects in Ref.\cite{teo} by considering a theory of embedding, which will be interesting for future studies.

In addition to the naturally developed atomic line defects during growth, the embedded quantum structures can be created and/or manipulated post-growth by atomic defect engineering using modern techniques such as atomic force microscope \cite{afm}, STM \cite{stm}, and electron beam lithography \cite{ebl1,ebl2}. Patterning the atomic line defects can in principle produce more desirable structures, such as the T-junctions, for studying the physics of MZMs. These findings open up new possibilities toward materializing quasi-1D topological superconductors with Majorana end states at high operating temperatures, which are essential for fault-tolerant quantum computing.

\section{ACKNOWLEDGMENTS}

We thank Jiangping Hu, Cheng Chen, and Yue Yu for useful discussions. Y.Z. and F.C.Z. are supported in part by the strategic priority research program of the Chinese Academy of Sciences Grant No. XDB28000000 and NSF China Grant No. 11674278. Y.Z. is also supported in part by NSF China Grant No. 12004383 and the Fundamental Research Funds for the Central Universities.  J.W. is supported by the National Key Research and Development Program of China (2018YFA0305604 and 2017YFA0303300) and the NSF China Grant No. 11888101 and Grant No. 11774008.  Z.W. is supported by the U.S. Department of Energy, Basic Energy Sciences Grant No. DE-FG02-99ER45747. K.J. acknowledges support from the start-up grant of IOP-CAS.

\appendix

\section{Calculation of the topological invairants}
In this appendix, we provide more detailed calculations of the topological invariants in the effective 1D models, including the $Z_2$ invariant in the time-reversal invariant class $DIII$ and the time-reversal symmetry breaking class $D$ \cite{altland,qi10,ardonne,ardonne2}, as well as the topological $Z$ invariant in the chiral symmetric classes $AIII$ and $BDI$. The general form of the mixed parity pairing order parameter is given by
\begin{equation}
\mathbf{\Delta}(k)=\Delta_s(k)i\sigma_y + \boldsymbol{d}(k)\cdot{\boldsymbol\sigma}(i\sigma_y).
\end{equation}
For nearest neighbor pairing, we have 	$\Delta_s(k)=2\Delta_s \cos k$
and $\vert \boldsymbol{d}(k)\vert=\Delta_t(k)=2\Delta_p \sin k$.

\subsection{Time-reversal symmetric case}
In the effective 1D model in Eq.~(4), the triplet-pairing $\boldsymbol{d}$-vector points along the $y$-direction, i.e.
$\boldsymbol{d}(k)=d_y(k){\hat y}$ and $\boldsymbol{d}(k)\cdot{\boldsymbol\sigma}=d_y(k)\sigma_y$.
The pairing order parameter is thus given by,
\begin{eqnarray}
	\Delta_{k}&=&i\sigma_y \Delta_s(k)+i\sigma_0\Delta_t(k)
\end{eqnarray}
In the presence of time-reversal symmetry, Ref.~[\onlinecite{qi10}] introduced a simple equation for calculating the topological $Z_2$ invariant ${\cal N}$ from the time-reversed pairing functions for each band $n$ at momentum $k$,
\begin{eqnarray}
	\delta_{nk}=\langle n,k| {\cal T} \Delta_k^{\dagger} |n,k \rangle
\end{eqnarray}
where $|n,k \rangle$ denotes the eigenstate and ${\cal T}=i\sigma_y{\cal K}$ is the time-reversal operator. In the weak pairing limit, the topological invariant ${\cal N}$ in 1D is given by the product
\begin{eqnarray}
	{\cal N}=\Pi_{s} [{\rm sgn}(\delta_s)]
\end{eqnarray}
where $s$ runs over all Fermi points of all bands between $0$ and $\pi$.
For the effective 1D model in Eq.~(4) of the main text, we thus obtain
\begin{eqnarray}
	{\cal N}={\rm sgn}[(\Delta_{s}(k_+) + \Delta_{t}(k_+))\times \nonumber \\
(\Delta_{s}(k_{-})-\Delta_{t}(k_{-}))]
\label{n}
\end{eqnarray}
where $k_{\pm}$ are the momenta of the two Fermi points between $k=0$ and $k=\pi$ of the Rashba-split bands. This simplifies to Eq.~(7) in the main text for nearest neighbor mixed parity pairing. A nontrivial invariant (${\cal N}=-1$) corresponds to a 1D time-reversal invariant TSC in class $DIII$.

\subsection{Time-reversal symmetry breaking case}
When the time-reversal symmetry is broken, the TSC in class $D$ is described by a different nontrivial $Z_2$ topological invariant in 1D, which can be calculated in terms of the Pfaffians of the BdG Hamiltonian in the Majorana basis at $k=0$ and $k=\pi$ \cite{ardonne2,jian}.
We thus need to rewrite the BdG Hamiltonian in terms of the Majorana fermion operators
\begin{eqnarray}
	\gamma_{i,a\sigma}&=&f_{i,\sigma}+f_{i,\sigma}^\dagger \\
	\gamma_{i,b\sigma}&=&-i(f_{i,\sigma}-f_{i,\sigma}^\dagger)
\end{eqnarray}
where $\sigma=\uparrow,\downarrow$ denotes each spin sector. For simplicity, consider the case where the time-reversal symmetry is broken by the Zeeman field $h_z$ in Eq.~(16) in the main text. The total Hamiltonian $H=H_{1D}+h_z\tau_z\sigma_z$ written in terms of Majorana operators in Fourier space is given by
\begin{eqnarray}
	H=\frac{i}{2}\sum_{k} \gamma_{k}^\dagger \hat{A}_k \gamma_{k},
\end{eqnarray}
where $\gamma_k=(\gamma_{k,a\uparrow},\gamma_{k,a\downarrow},\gamma_{k,b\uparrow},\gamma_{k,b\downarrow})^{T}$
\begin{widetext}
\begin{equation}
  \hat{A}_k=\begin{bmatrix}
		0 & 0 & \varepsilon_L(k)+i\Delta_t(k)+h_z & -2i\alpha_R\sin{k}+\Delta_s(k)\\
		0 & 0 & 2i\alpha_R\sin{k}-\Delta_s(k) & \varepsilon_L(k)+i\Delta_t(k)-h_z \\
		-\varepsilon_L(k)+i\Delta_t(k)-h_z  & 2i\alpha_R\sin{k}+\Delta_s(k) & 0 & 0 \\
		-2i\alpha_R\sin{k}-\Delta_s(k) & -\varepsilon_L(k)+i\Delta_t(k)+h_z  & 0 & 0
	\end{bmatrix}
\end{equation}
\end{widetext}
The 1D topological $Z_2$ invariant in class D is obtained as \cite{ardonne2,jian}
\begin{eqnarray}
	{\cal N}_{D}={\rm sgn}[{\rm Pf}(\hat{A}_{k=0}){\rm Pf}(\hat{A}_{k=\pi})].
\end{eqnarray}
Calculating the Pfaffian ${\rm Pf}(\hat{A_k})$ at $k=0$ and $k=\pi$, one obtains,
\begin{eqnarray}
	{\cal N}_{D}={\rm sgn}[(h_z^2-(\mu+2t)^2-4\Delta_s^2) \times \nonumber\\
	(h_z^2-(\mu-2t)^2-4\Delta_s^2)],
	\label{nh}
\end{eqnarray}
where a nearest neighbor $\Delta_s(k)=2\Delta_s\cos k$ is used. A nontrivial ${{\cal N}_D=-1}$ corresponds to a TSC in class $D$. Note that the odd-parity pairing $\Delta_t(k)$ does not enter the $Z_2$ invariant. Eq.~(\ref{nh}) shows that a time-reversal symmetry breaking TSC in class $D$ requires a large enough Zeeman field satisfying the condition {$h_z^2\in[(|\mu|-2t)^2+4\Delta_s^2,(|\mu|+2t)^2+4\Delta_s^2]$}.
As a result, the TSC in class $D$ is difficult to arise if the incipient impurity band is nearly particle-hole symmetric ($\mu\sim0$), but more easily to emerge when $\mu$ is close to the band top or band bottom. Such a 1D TSC supports a single MZM at each end of the RALD.
The above equation can be used to calculate the topological invariant ${\cal N_D}$ for Eq.~(11) in the main text.
In this case, Eq.~(11) in the Majorana basis becomes
\begin{equation}
 \hat{A}^{\pm}_k=\begin{bmatrix}
		i\Delta_t(k)\pm2i\alpha_R\sin{k} & -\varepsilon_L(k) \\
		\varepsilon_L(k) & -i\Delta_t(k)\pm2i\alpha_R\sin{k}
	\end{bmatrix}
\end{equation}
so that ${\rm Pf}(\hat{A}^{\pm}_k)=-\varepsilon_L(k)$ which gives the topological invariant for Eq.~(11) as
\begin{equation}
 {\cal N_D}={\rm sgn}[(2t+\mu)(2t-\mu)]
\end{equation}

\subsection{Chiral symmetric case}
The Hamiltonian $H(k)$ with the chiral symmetry can be block off-diagonalized by the unitary transformation $U$ that diagonalizes the chiral operator, so that we have
\begin{equation}
 UH(k)U^{\dagger}=\begin{bmatrix}
		0 & A(k) \\
		A^{\dagger}(k) & 0
	\end{bmatrix}
\label{offdiag}
\end{equation}
Since Det[$UH(k)U^{\dagger}$]=$\left|\text{Det}[A(k)]\right|^2$, Det[$A(k)$] will not vanish so long as the system is fully gapped. In this case, we can define a unit module complex function $z(k)=\text{Det}[A(k)]/\left|\text{Det}[A(k)]\right|$, whose winding number defines the topological invariant $Z$~\cite{tewari-sau}
\begin{equation}
 Z=\frac{-i}{2\pi}\int_{k=0}^{k=2\pi} \frac{dz(k)}{z(k)}.
 \label{winding}
\end{equation}
This formula works for both class $AIII$ and class $BDI$ discussed in the main text.
For instance, the Hamiltonians $H_A^\pm$ in Eq.~(5) in the main text describes two 1D models in class $AIII$ in the mirror eigen space can be brought to off-diagonal forms by the $2\times2$ unitary transformation $U_1=e^{-\frac{i\pi}{4}\tau_y}$ as
\begin{equation}
 U_1H_{A}^{\pm}U_1^{\dagger}=[\varepsilon_L(k)\mp 2\alpha_R \sin k] \tau_x + [\Delta_t(k) \mp \Delta_s(k)]\tau_y.
 \nonumber
\end{equation}
The off-diagonal blocks in Eq.~(\ref{offdiag}) are thus complex scalars,
\begin{equation}
 A^{\pm}(k)=\varepsilon_L(k)\mp 2\alpha_R \sin k-i[\Delta_t(k) \mp \Delta_s(k)]. \end{equation}
Eq.~(\ref{winding}) then leads two nontrivial winding numbers $Z^{\pm}=1$ in the two mirror eigen spaces, guaranteeing the presence of a pair of zero-energy bound states at each end of the mirror-line RALD.

As discussed in the main text, when the mirror-line RALD described by Eq.~(4) with a vanishing $\alpha_D$ is placed in a Zeeman field in the $xz$-plane, the corresponding 1D model
\begin{eqnarray}
H_{1D}(k)&=&\varepsilon_L(k)\tau_z\sigma_0+2\alpha_R \sin k \  \tau_z \sigma_y+\Delta_s(k) \tau_y \sigma_y
\nonumber \\
&+& \Delta_t(k) \tau_y\sigma_0+h_x\tau_z\sigma_x+h_z\tau_z\sigma_z
\end{eqnarray}
describes a superconductor in the topological class $BDI$. This Hamiltonian can be block off-diagonalized by the $4\times4$ unitary transformation $U_2=e^{-\frac{i\pi}{4}\tau_y\sigma_0}$,
\begin{eqnarray}
U_2H_{1D}(k)U_2^{\dagger}&=&\varepsilon_L(k)\tau_x\sigma_0+2\alpha_R \sin k \  \tau_x \sigma_y+\Delta_s(k) \tau_y \sigma_y \nonumber
\\
& + &\Delta_t(k) \tau_y\sigma_0+h_x\tau_x\sigma_x+h_z\tau_x\sigma_z.
\nonumber
\end{eqnarray}
The upper off-diagonal block in Eq.~(\ref{offdiag}) becomes
\begin{eqnarray}
A(k)&=&[\varepsilon_L(k) - i \Delta_t(k)]\sigma_0+[2\alpha_R \sin k -i\Delta_s(k)]\sigma_y \nonumber \\
 &+&h_x\sigma_x+h_z\sigma_z,
\end{eqnarray}
and its determinant is given by
\begin{eqnarray}
 {\rm Det}[A(k)]&=&[\varepsilon_L(k)-i \Delta_t(k)]^2+[\Delta_s(k)+2i\alpha_R \sin k]^2
\nonumber  \\
 &-&h_x^2-h_z^2.
\end{eqnarray}
The topological invariant $Z$ can be obtained by calculating the winding number using Eq.~(\ref{winding}), which evolves from $Z=2$ to $Z=1$ with increasing Zeeman field and eventually becomes trivial with $Z=0$ for large enough Zeeman field.

Following the discussion in the main text, there exists a hidden chiral symmetry ${\cal\widetilde C}$ associated with the anomalous time reversal reversal operator ${\cal\widetilde T}$ obeying ${\cal\widetilde T}^2=+1$ in the 1D model, despite the broken TRI by $h_z\neq0$. Thus for $h_z$ not large enough to satisfy Eq.~(\ref{nh}) and the system has a trivial invariant for class $D$, but the RALD remains a TSC in class $BDI$ characterized by an integer $Z$ topological invariant \cite{tewari-sau,tewari} with a pair of MZMs at each end.
Increasing $h_z$ eventually leads to topological transitions in class $BDI$ from $Z=2$ to $Z=1$ with a single MZM at each end of the RALD.

\end{document}